\documentclass[manuscript]{aastex}

\newcommand{\dxdy}[2]{{\frac{\partial{#1}}{\partial{#2}}}}

\newcommand{\DxDy}

\shorttitle{Toroidal field reversals}
\shortauthors{Rogers}

\begin{document}


\title{Toroidal Field Reversals and the axisymmetric Tayler Instability}
  

\author{T.M. Rogers}
\affil{Department of Planetary Sciences, University of Arizona,
    Tucson, AZ, 85719}




\begin{abstract}
We present axisymmetric numerical simulations of the solar interior, including the convection zone and an extended radiative interior.  We find that differential rotation in the convection zone induces a toroidal field from an initially purely poloidal field.  This toroidal field becomes unstable to the axisymmetric Tayler instability and undergoes equator-ward propagating toroidal field reversals. These reversals occur in the absence of a dynamo and without accompanying poloidal field reversals.  The nature and presence of such reversals depends sensitively on the initial poloidal field strength imposed, with North-South symmetric reversals only seen at a particular initial field strength.  Coupled with a dynamo mechanism which regenerates the poloidal field this could be one ingredient in the sunspot cycle.  
\end{abstract}





\section{Introduction}

One of the fundamental questions in solar physics is how the Sun generates its magnetic field.  The most likely scenario is that the large-scale solar magnetic field is maintained by a hydromagnetic dynamo.  This is supported by the remarkable regularity of the solar magnetic cycle.  Over the course of the sunspot cycle bipolar pairs of magnetic regions appear at mid-latitudes over time, with new pairs appearing at lower and lower latitudes.  These pairs have a distinct tilt in latitude (Joy's Law), with the leading sunspot (the one of the pair closest to the equator) having one polarity and the trailing sunspot having the opposite polarity.  In the other hemisphere the polarities are reversed (Hale's law), so that the leading sunspots in the Northern hemisphere have the opposite sign of the leading sunspots in the Southern hemisphere.  These sunspot pairs are associated with toroidal magentic field erupting to the surface, explaining their bi-polar nature.  The leading sunspot polarity has the same polarity as the large scale poloidal field in that hemisphere and, after approximately 11 years, the sunspot cycle repeats, with the large scale magnetic polarity as well as that of the leading sunspot polarity reversing in each hemisphere reversed.  The sunspot cycle itself is then said to have a period of approximately 11 years, while the global magnetic cycle has a period of approximately 22 years.   

Exactly how this dynamo process works, and produces these detailed observations of the sunspot and magnetic cycle remains the subject of intense research.  Dynamo theorists usually envision a process whereby a large scale poloidal field is acted upon by differential rotation (most likely in the region of strongest shear, the tachocline), generating a strong toroidal field which then becomes magnetically buoyant and rises through the convection zone to produce sunspot pairs at the solar surface.  Most research done on the solar dynamo has concentrated on solving mean field equations \citep{Steenbeck:1966p2562,Moffatt:1972p2567}, in which the induction of field due to correlations of fluctuating velocity and magnetic field are prescribed by the so-called $\alpha$ effect and the differential rotation prescribed by the $\Omega$ effect.  These reduced models, with input from observations, can reproduce some of the features of the solar magnetic and sunspot cycle in the form of reversing toroidal fields which propagate toward the equator \citep{Charblr,Dikpati:2008p2906}, albeit with tunable parameters. However, it has been more difficult to produce equatorward propagation of toroidal field in more realistic simulations which solve the full magnetohydrodynamics (MHD) equations \citep{Glatzmaier:1985p2917,Glatzmaier:1985p2914,Brun:2004p2818,Brown:2010p2752}.  Early simulations \citep{Glatzmaier:1985p2917} could produce a dynamo with oscillating toroidal field, but the period was usually much shorter than the 11 year sunspot cycle and toroidal field tended to propagate toward the pole.  More recent simulations \citep{Brown:2010p2752} have also been able to robustly produce dynamos, some of which show reversing toroidal field polarity, but again, the period is not solar like and are generally poleward propagating or stationary \citep{Brown:2010p2752,Ghizaru:2010p2922}.  So it appears from these simulations that dynamos are robust, but dynamos which produce solar-like properties, such as equator-ward propagation of toroidal field are substantially more difficult to produce.  

While a hydromagnetic dynamo is the most likely explanation of the solar cycle, a magnetic oscillator is not completely ruled out.  The dissipation time of a primordial large scale magnetic field in the Sun is $\approx 10^{10}$ years, therefore a decaying field superposed by an oscillating flow could possibly reproduce some of the solar observations.  

Most of the previous numerical simulations of dynamo action in the Sun and stars have focused on the convection zone, where it is thought that the rise and twist of magnetic field could produce a robust $\alpha$ effect.  \cite{Spruit:1999p1008,Spruit:2002p1023} has argued that a dynamo could be driven in stellar radiative interiors requiring only differential rotation.  In his picture the $\alpha$ effect could be generated by the Tayler instability.  \cite{Braithwaite:2006p110} followed up this analytic work with numerical simulations showing that a dynamo could be generated in stellar interiors, requiring only some small differential rotation.  However, \cite{Brun:2006p172} conducted similar numerical simulations of the solar radiative interior and found the instability but no regeneration of the mean poloidal field and therefore, concluded that no dynamo was present.  Regardless of the presence of a dynamo, the presence of the Tayler instability in radiative interiors appears to be robust.     

The simulations presented here are axisymmetric and therefore, there is no dynamo.  However, the simulations do reproduce some interesting features; namely, equatorward propagation of reversing toroidal field, in the absence of a dynamo or a reversing poloidal field.  The rest of the paper is organized as follows.  In section 2 we discuss the numerical model, in section 3 we discuss the evolution of the toroidal field, the role of the Tayler instability and the effect of the initial field strength.  In section 4 we conclude with a discussion of the general applicability of this model to stellar interiors given its admittedly crude set up.  

\section{Numerical Model}

We solve the full magnetohydrodynamic (MHD) equations in axisymmetric, spherical coordinates in the anelastic approximation.  The numerical scheme is a combination of the equations outlined in \cite{Rogers:2005p1533} with the numerical scheme of \cite{Glatzmaier:1984p2768}.  The most substantial differences between this model and the \cite{Glatzmaier:1984p2768} model are that here we use a finite difference scheme in radius, our model is axisymmetric, and we solve the temperature equation rather than an entropy equation.  For more details on the numerical method, see Appendix A.

\subsection{Model Setup}

We solve the MHD equations in axisymmetric spherical coordinates in the radial range from 0.15 R$_{\odot}$ to 0.95 R$_{\odot}$.  The radiative region extends from 0.15 R$_{\odot}$ to 0.71 R$_{\odot}$ and the convection zone extends from 0.71 R$_{\odot}$ to 0.90 R$_{\odot}$ and we impose an additional stably stratified region from 0.90 R$_{\odot}$ to 0.95 R$_{\odot}$.  The reference state thermodynamic variables are taken from a polynomial fit to the standard solar model from \cite{ChristensenDalsgaard:1996p2948}. Therefore, the density ranges from approximately 60 $gm/cm^{3}$ at the bottom of the radiation zone to $10^{-2} gm/cm^{3}$ at the top of the domain.  The stable stratification is that given by the standard solar model and the super-adiabaticity of the convection zone is set to $10^{-7}$.  The grid resolution is 512 latitudinal grid points by 1500 radial grid points, with 500 radial zones dedicated to the radiative interior and 1000 zones dedicated to the tachocline and convection zone.  

Because this model is not fully three-dimensional (3D) we can not reproduce the Reynolds stresses required to set up the observed differential rotation \citep{thomp03} in the convection zone and tachocline.  We therefore, impose this differential rotation through a forcing term on the azimuthal component of the momentum equation:
\begin{eqnarray}
\dxdy{}{t}(\frac{U_{\phi}}{r \sin\theta})&=&-\frac{(\nabla\cdot{\bf u}{U_{\phi}})_{\phi}}{r\sin\theta}+\frac{((\nabla\times{\bf B})\times{\bf B})_{\phi}}{\bar{\rho}\mu r\sin\theta}+\frac{(2{\bf u}\times\Omega)_{\phi}}{r\sin\theta}\nonumber\\&
&{+\frac{\nu}{r\sin\theta}\nabla^{2}u_{\phi}+\frac{\nu u_{\phi}}{r^{3}\sin^{3}\theta}+\frac{1}{\tau}\left(\Omega'(r,\theta)-\frac{u_{\phi}(r,\theta)}{r\sin\theta}\right)}
\end{eqnarray}
where the last term on the rhs represents the forcing term.  $\Omega'(r,\theta)$ is the prescribed (observed) differential rotation relative to the constant $\Omega$ and $\tau$ is the e-folding time of the forcing which we set to $10^{4}$.\footnote{We ran models with two other values of $\tau$ ($10^3$ and $10^{5}$) for a fraction of the total time of the models ran here and saw little difference in the meridional circulation or azimuthal flow variation.}.  The rotation profile is given by
\begin{equation}
\begin{array}{rclr}
\Omega'(r,\theta)&=&\Omega & \mbox{for $r < 0.67R_{\odot}$}\\
\Omega'(r,\theta)&=&(A+B\cos^{2}\theta+C\cos^{4}\theta-\Omega){\bf erf}(\frac{2(r-0.67R_{\odot})}{0.05R_{\odot}})  &\mbox{for $0.67<r<0.72$}\\
\Omega'(r,\theta)&=&(A+B\cos^{2}\theta+C\cos^{4}\theta-\Omega) & \mbox{for $r>0.72$}\\
\end{array}
\end{equation}
where $\Omega$ is the constant value set to 441nHz, A is 456nHz, B is -42nHz and C is -72 nHz \citep{thomp03}.

\begin{figure}

\centering
\includegraphics[width=6in]{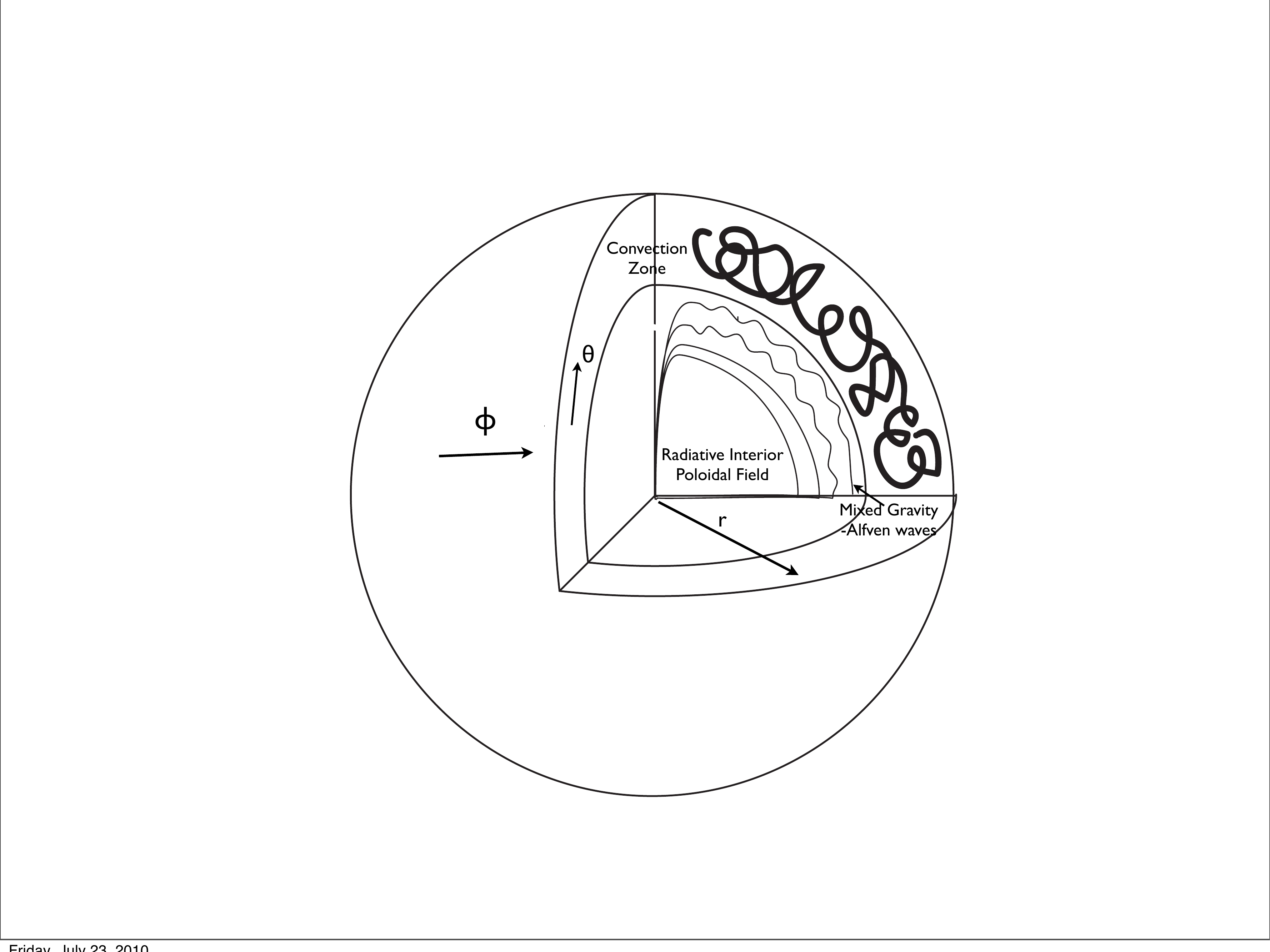}
\caption{Model schematic.  Radiative interior extends from 0.15R$_{\odot}$ to 0.71 R$_{\odot}$ and convection zone lying from 0.71R$_{\odot}$ to 0.90R$_{\odot}$.  After convection is established a dipolar field is imposed in the radiative interior.  Waves of mixed gravity-Alfven type exist in the radiative interior.}

\label{fig:setup}
\end{figure}

\begin{figure}

\includegraphics[width=4in]{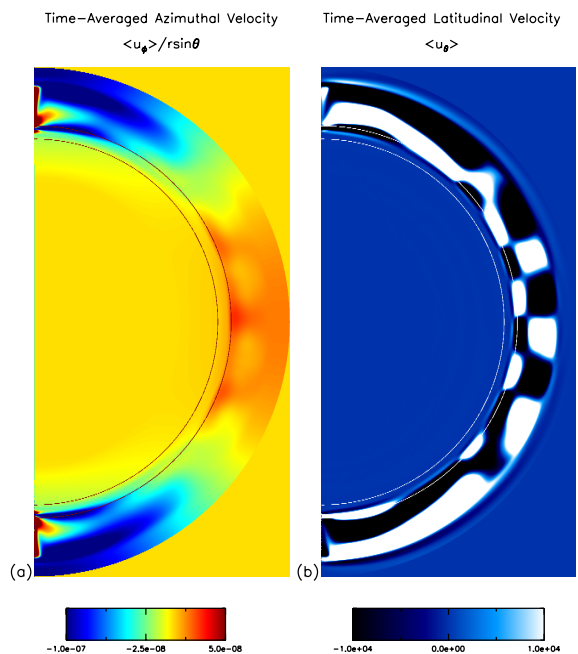}
\caption{Time averaged differential rotation for Model B is shown in (a) with red representing flows faster than the mean and blue representing flows slower than the mean.  Red overlaid lines represent the extent of the initially imposed tachocline.  The figure shows the forced differential rotation as described in the text, plus any fluctuations averaged in time.  The amplitudes in the color bar are in Hz.  Time averaged meridional circulation for Model B (b), white colors represent flow towards the south pole, while dark colors represent flow towards the north pole.  One can then clearly see poleward flow at the top of the convection zone, and equator-ward flow at the base of the convection zone.  Amplitudes in the color bar are quoted in cm/s.  White overlaid lines represent the extent of the initially imposed tachocline. Both time-averages are over the entire simulation ($2\times 10^{8}s$).}
\label{fig:azlatvel}
\end{figure}

Initially, we run a purely hydrodynamic model until convection is sufficiently established at which point we imposed a dipolar field with the form:

\begin{equation}
B_{1}=B_{o}r^{2}(1.0-\frac{r}{0.71R_{\odot}})^{2}
\end{equation}
so that initially all field lines close within the radiative interior, but overlap the bottom of the tachocline.  We ran five models, which we label Models A,B,C,D, and E with initial field strengths, $B_{o}$, of 10G, 40G, 80G, 200G and 4000G, respectively, with all other parameters the same.  Figure~\ref{fig:setup} shows a schematic of the initial setup and model, while Figure~\ref{fig:azlatvel} shows the time-averaged azimuthal velocity as forced in Equation (2), along with the time-averaged meridional flow that results from this differential rotation profile (b).

\section{Results}

\subsection{Toroidal Field Evolution}

\begin{figure}
\includegraphics[width=4in]{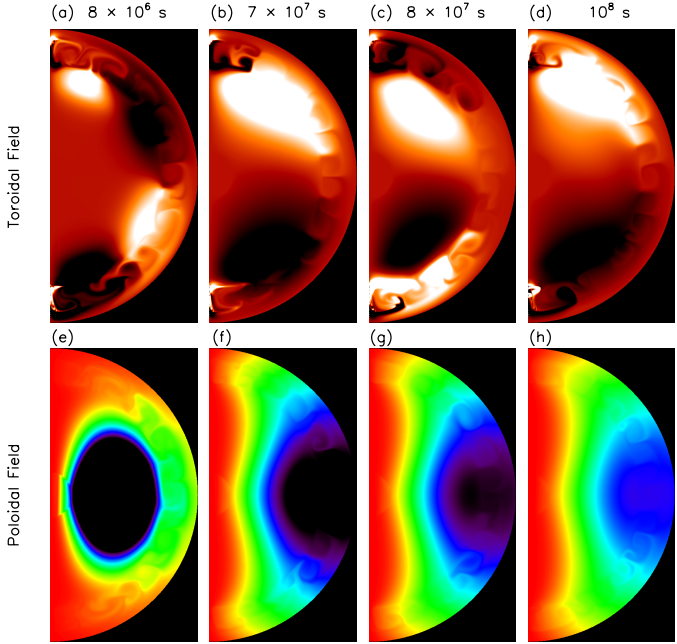}
\caption{Time snapshots of the toroidal field (a-d) and poloidal field (e-h).  In (a-d) white represents positive toroidal field, while black represents negative.  Initially, the radial gradient of the differential rotation dominates the induction of toroidal field, this as seen as two signs of toroidal field in each hemisphere (a).  However, quickly the advection of toroidal field by the meridional flow becomes the dominant term controlling toroidal field evolution and one sign in each hemisphere remains.  A toroidal field reversal in the tachocline and convection zone is seen between (b) and (c), see corresponding time in Figure~\ref{fig:reversaladdb}.  Poloidal field is seen in (e-h) and is diffusing outward and decaying in time.  The colormaps in the various images are slightly different because of the steady increase in toroidal field strength and steady decrease in poloidal field strength.  Peak toroidal field in this model is approximately $\pm 50G$.  Both poloidal and toroidal field effectively connect to the convection zone despite the meridional circulation induced by the differential rotation. }
\label{fig:fieldevol}
\end{figure}

In the following we discuss the primary results from Model B ($B_{o}=40G$) and leave the discussion of the variation in field strength for Section 3.3.  Because of the slight overlap of the imposed dipolar poloidal field with the tachocline and because of magnetic diffusion, it does not take long for the dipolar poloidal field to be stretched into toroidal field by the differential rotation in the tachocline and convection zone.  Initially, this induces a toroidal field which has two signs each in the Northern and Southern hemisphere, see Figure~\ref{fig:fieldevol}.  This structure can be understood by looking at the azimuthal component of the magnetic induction equation:

\begin{eqnarray}
\dxdy{B_{\phi}}{t}&=&rB_{r}\dxdy{}{r}(\frac{u_{\phi}}{r})-ru_{r}\dxdy{}{r}(\frac{B_{\phi}}{r})+\frac{\sin\theta B_{\theta}}{r}\dxdy{}{\theta}(\frac{u_{\phi}}{\sin\theta})-\frac{\sin\theta u_{\theta}}{r}\dxdy{}{\theta}(\frac{B_{\phi}}{\sin\theta}) \nonumber\\&
&\mbox{}+B_{\phi}u_{r}h_{\rho}+\eta\nabla^{2}B_{\phi}-\eta\frac{B_{\phi}}{r^{2}\sin^{2}\theta}
\end{eqnarray}
The structure of the differential rotation gives $\partial{u_{\phi}}/\partial{r}$ negative at high latitudes and positive at low latitudes and the first term on the rhs of (4) induces oppositely signed toroidal field at high and low latitudes, with asymmetry about the equator.  This is seen in Figure~\ref{fig:fieldevol}a.  

However, the first term on the rhs of (4), induction due to radial gradient in the rotation rate, is dominant for only a short time.  The latitudinal toroidal field gradient becomes unstable (see Figure~\ref{fig:compositeaddb}, section 3.2) and the dominant induction term becomes the fourth term on the rhs of (4), advection of toroidal field by latitudinal flow.  This leads to a single dominant signed toroidal field in each hemisphere, asymmetric about the equator.  In the following I will refer to positive toroidal field (white in figure 3) as the dominant sign in the Northern hemisphere, and negative toroidal field (black in figure 3) as the minority field sign in the Northern hemisphere and conversely, in the Southern hemisphere.  These particular signs arise naturally from the equator-ward directed $u_{\theta}$ coupled with a toroidal field which is stronger at the poles than the equator.  Subsequently, and for most of the simulation, the advection of toroidal field by meridional flow (fourth term on the rhs of equation 4) is dominant within the tachocline (figure 3b).  As can be seen from Figure~\ref{fig:fieldevol}b to Figure~\ref{fig:fieldevol}c a ``reversal'' takes place in which the dominant field is replaced by the minority field in the bulk of the convection zone and tachocline.  This reversal proceeds by the advection of minority signed field toward the equator by meridional circulation at the base of the convection zone.  

\begin{figure}
\includegraphics[width=6in]{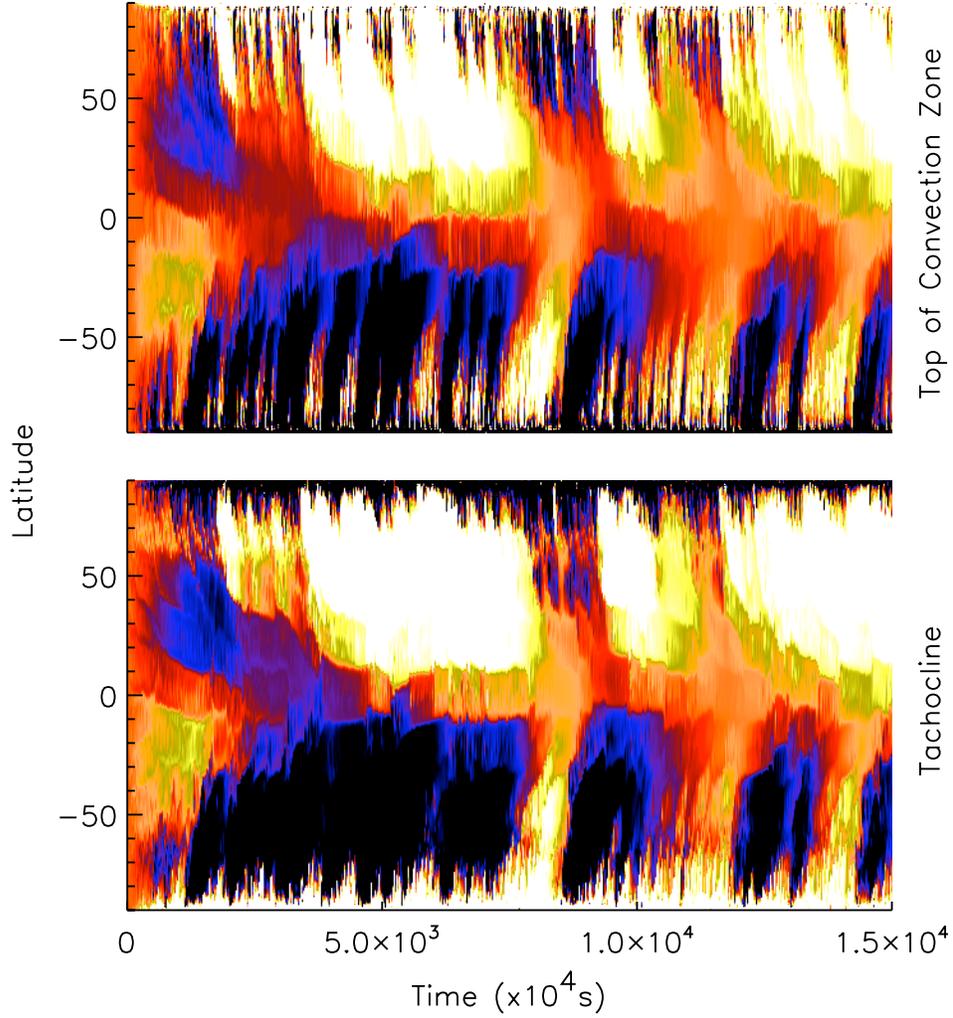}
\caption{Toroidal magnetic field as a function of time and latitude, in the tachocline (bottom) and at the top of the convection zone (top).  Toroidal field reversals with equator-ward propagation are seen.  The Northern hemisphere shows a weaker second reversal and no third reversal.}
\label{fig:reversaladdb}
\end{figure}

The evolution in time of the reversal process is better demonstrated in a time-lattitude plot, as seen in Figure~\ref{fig:reversaladdb}a.  Shown there is the toroidal field amplitude as a function of time and latitude.  One can see the initial phase of two signs in each hemisphere giving way to a dominant sign in each hemisphere.  Reversals subsequent to that shown in Figure~\ref{fig:fieldevol} are also seen here.  During the first reversal (around $7\times 10^{7}$s, or around 30 rotation periods) the northern and southern hemispheres are relatively in phase, however later reversals are stronger in the southern hemisphere and the coherence between north and south is not strictly maintained.  As is seen in Figure~\ref{fig:reversaladdb}, these reversals are seen both in the tachocline and at the top of the convection zone.  We note that no reversal occurs in the radiative interior, nor does any reversal occur in the poloidal field.  
 
Except for the very initial phases the dominant term in the tachocline responsible for toroidal field evolution is the advection of toroidal field by the meridional circulation.  The time averaged equator-ward latitudinal velocity at the base of the convection zone is $\approx 3 \times 10^{3} cm/s$, which would imply an advection time from pole to equator of $\approx 3 \times 10^{7} s$ ($\approx$ 30 rotation periods).  Looking at Figure~\ref{fig:reversaladdb} one can determine an approximate time for toroidal field to propagate from the pole to the equator to be $\approx 1.5 \times 10^{7}-2\times 10^{7} s$ during a reversal.  This slight discrepancy is likely due to the difference between the time-averaged latitudinal velocity compared to the typical latitudinal velocity of the convective cells, which at the base of the convection zone is $\approx 10^{4}-10^{5} cm/s$.  This indicates that the bulk of the latitudinal transport is due to time-varying convection rather than the time-averaged meridional circulation.\footnote{Note that the convective velocities in this simulation are higher than that expected in the solar convection zone.  This is because our enhanced thermal diffusion leads to a larger heat flux through the system, requiring a larger convective velocity to carry it, given the same temperature gradient.  In these simulations these effect leads to an enhancement of the flux by $10^{6}$ and hence, convective velocities enhanced by $\approx$ 100.}  In animation of the images shown in Figure~\ref{fig:fieldevol}a-d one can see that the toroidal field is mostly advected by the convective eddies, therefore leading to a slightly more rapid equator-ward propagation than would be implied by the mean latitudinal flow.

\begin{figure}
\includegraphics[width=5in]{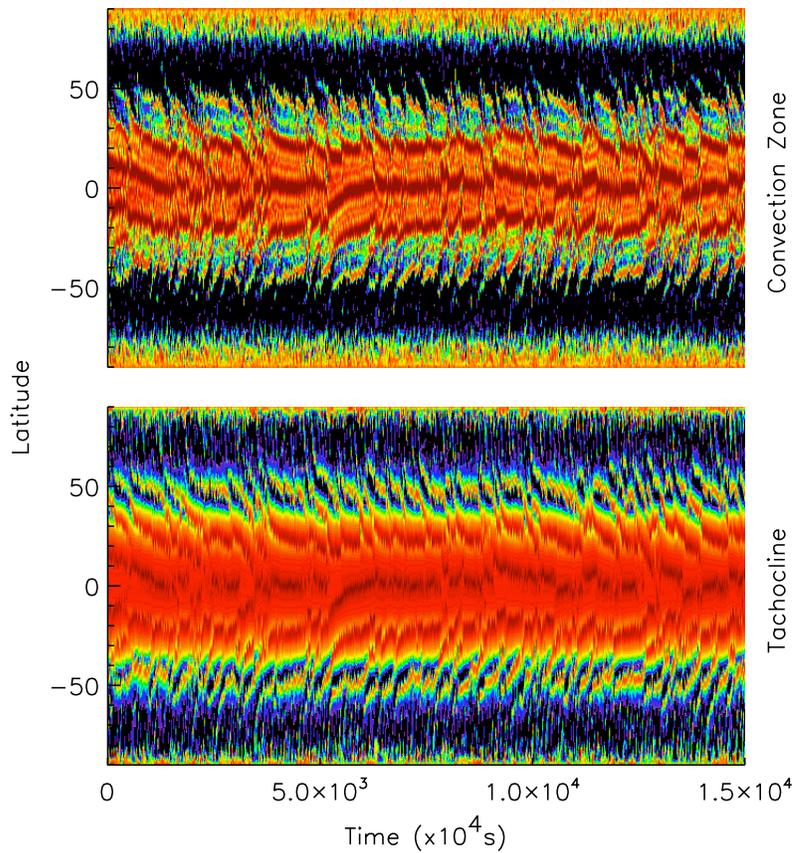}
\caption{Azimuthal velocity as a function of time and latitude, in the tachocline (bottom) and in the convection zone (top).  One can see some signs of short period torsional oscillations which propagate toward the equator in time.  These are symmetric about the equator and relatively periodic.  However, their timescale is more like that of convective eddies rather than magnetic field reversals, unlike the Sun.}
\label{fig:aztime}
\end{figure}

Over the duration of the simulation, the polodial field does little else than diffuse and be advected by the overlying convection zone, as can be seen in figure 3e-h.  As this model is axisymmetric there can clearly be no regeneration of the poloidal field and hence, no dynamo.  Therefore, one would expect the reversals of field sign to be due to fluctuations of the azimuthal velocity superimposed on the imposed differential rotation.  In Figure~\ref{fig:aztime} we show the azimuthal velocity as a function of time and latitude.  One can see the general behavior of the (imposed) observed differential rotation with fast equator (prograde flow) and slow poles (retrograde flow), this is maintained throughout the simulation.  However, one can also see weak signs of prograde flow at both poles, intermittent in the tachocline but persistent in the convection zone.  Likewise one can see bands of prograde azimuthal flows starting at mid-latitudes with slight propagation toward the equator, although on much shorter timescales than the magnetic reversals, contrary to observations.  This short timescale structure with equator-ward propagation is also seen in the Figures 6-8, in the Tayler instability panel and, at some level, in the ratio of toroidal to poloidal field indicating that there is some interplay between magnetism and azimuthal flow variations, we are currently pursuing this interaction.  

Because of the lack of obvious variation in the azimuthal flow on the same timescales of the magnetic field it appears that the reversals must be due to a local change in sign and amplification in toroidal field followed by advection.  This can happen if the persistent minority sign seen at high latitudes is amplified and becomes dominant in the region.\footnote{This may only be possible with prograde poles, which we see here but are not necessarily realistic.}  

\subsection{Tayler Instability and toroidal field evolution}

There has been significant discussion about the possible role of magnetic instabilities in stellar interiors \citep{Wright:1973p2277,Tayler:1975p2839,Acheson:1978p9,Spruit:1999p1008,Parfrey:2007p3032}.  Although it is generally believed that convective or turbulent motions provide the required $\alpha$ effect to regenerate poloidal field from toroidal field, it was suggested by \cite{Spruit:2002p1023}, that convection is not needed because this effect could be provided by an instability in the field itself.  He argued that the most relevant instability in stellar interiors was likely to be the Tayler instability.  \cite{Parfrey:2007p3032} argued that the diffusive magneto-rotational instability (MRI) was likely at high latitudes in the tachocline and that while this region overlaps with the region of Tayler instability the MRI is likely to be more dynamically important because of its exponential growth.  They argued that this instability could disrupt magnetic fields, thus explaining the lack of sunspots at high latitude.  

Here we find that the toroidal field reversals observed in Figure~\ref{fig:reversaladdb} are consistent with times when the toroidal field is unstable to the axisymmetric Tayler instability as described by the simple instability criterion provided in \cite{goos81} and \cite{Spruit:1999p1008}, namely:

\begin{equation}
\cos\theta \dxdy{}{\theta}\left(ln(\frac{B^{2}}{\sin^{2}\theta})\right )>0
\end{equation}

\begin{figure}
\includegraphics[width=5.5in]{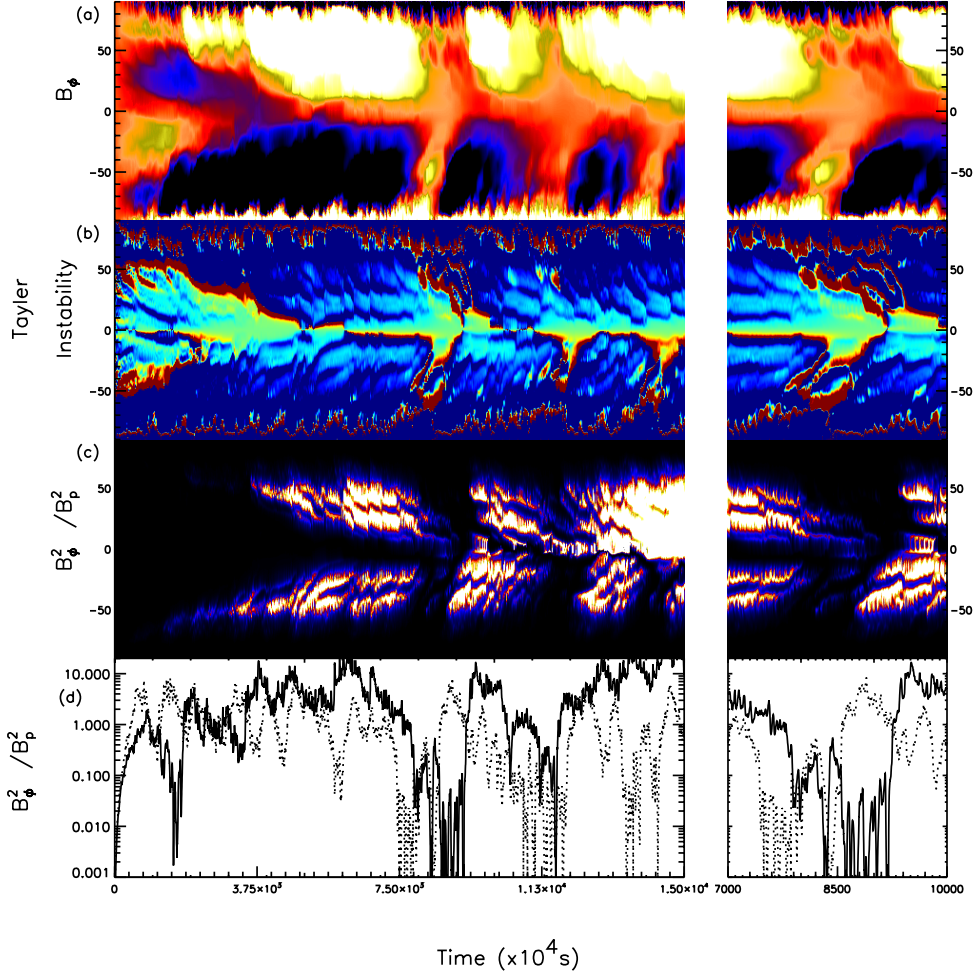}
\caption{Toroidal magnetic field as a function of time and latitude within the tachocline, with color scaled at $\pm 30G$ (a), Tayler instability criterion as described in Equation 5 with color scaled at $\pm 1$ (b), ratio of toroidal to poloidal magnetic field energies, with color scaled 0-300 and white representing high amplitude ratios (c) and ratio of toroidal to poloidal magnetic field energies at high latitudes (20$^{\circ}$) in the Northern hemisphere (solid line) and Southern hemisphere (dotted line) (d).  The right hand panels show the same values just zoomed over the first reversal period.}
\label{fig:compositeaddb}
\end{figure}

\begin{figure}
\includegraphics[width=6in]{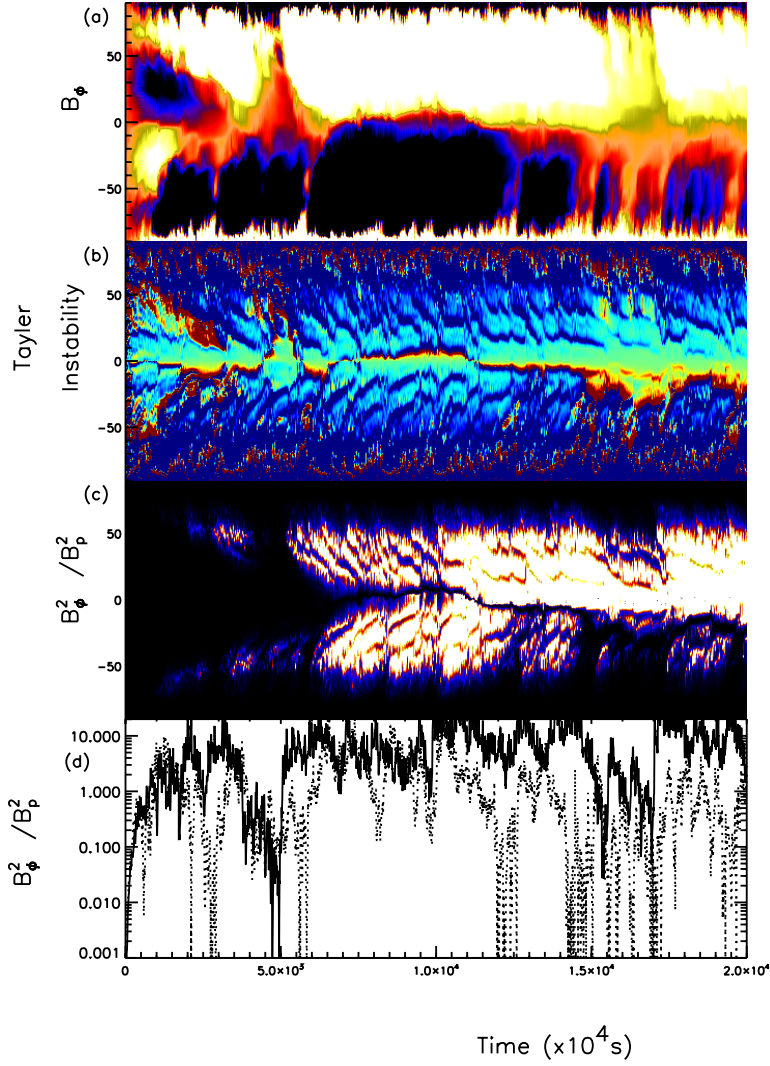}
\caption{Same as Figure~\ref{fig:compositeaddb}, but here showing Model E ($B_{o}=4kG$) and neglecting the zoomed panels.  Because this model has a larger initial field strength, (a) is scaled at $\pm 50G$, others are scaled the same as Figure~\ref{fig:compositeaddb}.}
\label{fig:compositeaddb3}
\end{figure}

\begin{figure}
\includegraphics[width=6in]{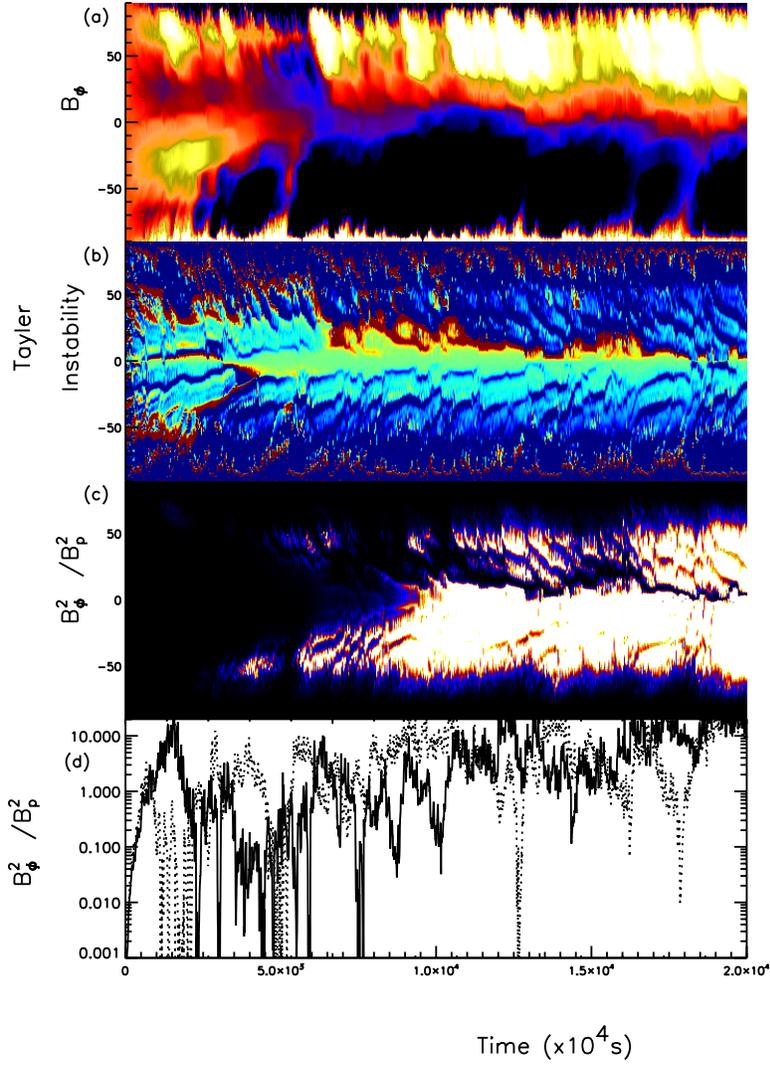}
\caption{Same as Figure~\ref{fig:compositeaddb}, but here showing Model C ($B_{o}=80G$).  Because this model has a weaker initial field strength, (a) is scaled at $\pm 3$.}
\label{fig:compositeaddb2}
\end{figure}

In Figure~\ref{fig:compositeaddb}b we show the lhs of the inequality of (5).  One can clearly see that reversals of the field are coincident with a Tayler instability that proceeds even at low latitudes.  The instability does not appear in the bulk of the radiative interior where the toroidal field strength is substantially lower due to weaker differential rotation and the (stabilizing) poloidal field is correspondingly stronger.  The instability is strongest in the tachocline and is present, but much less coherent in the convection zone.  The bulk of the time the instability is confined to higher latitudes as expected \citep{Parfrey:2007p3032}.  In Figure~\ref{fig:compositeaddb}d, which shows the ratio of the ${\it local}$ toroidal to poloidal field energy at high latitudes, one can see an almost oscillatory behavior of the instability.  At these high latitudes the dominant toroidal field grows due to differential rotation, once the local ratio reaches a value around 10-20, the instability disrupts the field growth and magnetic flux destruction ensues.  Normally, the field decay is weak and growth due to differential rotation is re-established shortly thereafter.  However, occasionally, field destruction caused by the instability is more severe and the toroidal field energy drops precipitously.  During these times the minority signed field can become dominant and be advected equator-ward.  Therefore, times of field reversals as seen in Figure~\ref{fig:compositeaddb}, ~\ref{fig:compositeaddb3} and ~\ref{fig:compositeaddb2}  are associated with minima of the toroidal field energy relative to the poloidal energy and these minima are induced by the Tayler instability.  

Indeed, one can see in Figure~\ref{fig:compositeaddb}d that the slight North-South asymmetry in reversal coincides with a slight discrepancy in timing between flux destruction in the Northern and Southern hemispheres.  This can be seen in all of the composite figures (6-8), when flux destruction occurs there is a reversal, when there is weak or no destruction, no reversal ensues.  

The stability condition expressed in (5) is for a purely toroidal field.  It has been shown \citep{Wright:1973p2277,Braithwaite:2004p2184} that mixed field configurations are more stable.  Therefore, in addition to the condition expressed in (5) the toroidal field must also overcome the stabilizing effect of the poloidal field.  This is what we see in the bottom panels of Figures~\ref{fig:compositeaddb}, ~\ref{fig:compositeaddb3} and ~\ref{fig:compositeaddb2}.  Substantial flux destruction only occurs when the criterion in (5) is met, along with the ratio of local toroidal to poloidal magnetic energy reaching some critical value.  The critical value appears to be around 10-20, but clearly is not strict and other factors, such as local turbulent dissipation could become relevant.  It should be noted that the non-axisymmetric Tayler instability and MRI are likely more relevant in real stars, this could lead to more regular flux destruction and possibly more regular reversals, although whether they would proceed like the ones here is unclear.  

\subsection{Dependence on Initial Field Strength} 

In the preceding section we concentrated on Model B and found that it showed equatorward propagating toroidal field reversals.  These reversals were precipitated by a Tayler instability and were sometimes coherent between Northern and Southern hemispheres, but not always.  We ran four additional models with varying initial field strengths.  All of the models, except Model A, show some evidence of reversal, however, only Model B shows reversals which are coherent in Northern and Southern hemispheres.  Furthermore, only Model B and E show reversals of any substantial duration.  Model A, which shows no reversal does show dominant toroidal field which weakens and then increases in time, but this is not accompanied by amplification and advection of minority field.

Figures 7 and 8 shows the toroidal magnetic field, Tayler instability criterion and ratio of toroidal to poloidal magnetic field energy for Model E and C, respectively.  One can see the initial two signed field structure giving way to a single dominant field in each hemisphere, similar to Model B.  Other equatorial propagting field reversals are also seen, particularly in the Southern hemisphere for Model E and the Northern hemisphere for Model C.  One should note that the local ratio of the toroidal to poloidal field strengths in both of these models is similar to that seen in Model B, despite the vast difference in initial poloidal field strength; again implying the ratio of the field strengths plays some role in determining the growth of the Tayler instability.  
It is unclear why Model B shows symmetric reversals, while the others do not, it is also not clear why the various reversals are so different both in latitudinal excursion and in duration.  These simulations indicate that weakening of the dominant toroidal field by the Tayler instability is robust, however, accompanying solar-like reversals may be possible only in limited parameter space of magnetic field strength.  The dynamics that dictate that parameter space are presently unclear.

\section{Discussion}
We have shown that a poloidal field, initially confined to the radiative interior and acted on by differential rotation of the kind observed in the Sun can lead to reversals and equator-ward propagation of toroidal field.  The process is initiated by differential rotation acting on an initially poloidal field to induce toroidal field.  When both the basic Tayler instability criterion expressed in (5) and the local ratio of toroidal to poloidal field energy reaches some critical value, instability and rapid flux destruction ensues.  During this time the minority field sign can become dominant and be transported toward the equator by meridional circulation.  Once the minority field decays and is advected away the dominant field remains and the process repeats. 

The relevance of this process in the Sun or any star are unknown.  First, the toroidal field reversals do not appear periodic, nor completely symmetric about the equator, although this depends on the initial field strength.  Second, the non-axisymmetric Tayler instability or the weak field MRI is likely dominant in the solar radiative interior and tachocline.  However, any instability that could lead to the amplification of the minority signed toroidal field could feasibly mimic what we see here.  In fact, a more robust instability may lead to more regular reversals.  Third, the high latitude fluctuations of azimuthal velocity which initiate this process, while out of sight of helioseismology, may be unrealistic.  Finally, this process clearly indicates a high-latitude branch of toroidal field which is equator-ward propagating, contrary to the observations.

While these are all serious issues which complicate the applicability of this model to the Sun, we think there are substantial relevant overlaps with the Sun that make this model interesting.  First, this is the first model which solves the full MHD equations with convection and rotation and gets equator symmetric equator-ward propagating and reversing toroidal field\footnote{Note, however, that \cite{mitra2011} have simulated equator-ward propagating reversing fields using forced MHD turbulence with imposed helicity.  Those models, implicitly including rotation and convection through helicity have produced much more regular reversals.} and, interestingly, it does so without the operation of a dynamo and without corresponding poloidal field reversals..  We also find that these reversals depend on the initially imposed field strength, and it is  possible that there is a preferred field strength which more adequately represents observations.  Of course, we would like to run more models to confirm this, but computational resources are limited.  Second, we have shown that the axisymmetric Tayler instability, likely to be operating in stellar interiors, can act to destroy the dominant toroidal field, allowing the weaker minority field to be advected equator-ward by the meridional circulation at the base of the convection zone.  This instability seems to act when the basic criterion expressed in (5) is met combined with when the local ratio of toroidal to poloidal field energies reaches some critical value.  Third, we have shown that meridional circulation can effectively act as a conveyor belt in advecting, at least in our 2D geometry, minority field toward the equator at the base of the convection zone; a requirement for many flux-transport dynamo models \citep{Charblr,Dikpati:2007p2879}.  
    
\bibliographystyle{apj}
\bibliography{field}

\acknowledgments
We are grateful to G. Glatzmaier, M. McIntyre and G. Ogilvie for helpful discussions.  We are extremely grateful to an anonymous referee for a meticulous report and thoughtful suggestions which have resulted in much improved manuscript. Support for this research was provided by a NASA grant NNX11AB46G.  T. Rogers is supported by an NSF ATM Faculty Position in Solar physics under award number 0457631.  Computing resources were provided by NAS at NASA Ames.  

\appendix
\section*{Appendix A: Numerical Method}

In this model we solve the full set of axisymmetric MHD equations in the anelastic approximation.  

\begin{equation}
\nabla\cdot(\overline{\rho}{\bf u})=0.
\end{equation}
\begin{equation}
\nabla\cdot{\bf B}=0
\end{equation}
\begin{eqnarray}
\dxdy{{\bf u}}{t}+({\bf u}\cdot\nabla){\bf u}&=&
-\nabla P - C\overline{g}\hat{r}+2({\bf u}\times\Omega)+\frac{1}{\bar{\rho}}({\bf J}\times{\bf B})\nonumber\\&     
&\mbox{}+\nu\left(\nabla^{2}{\bf u}+\frac{1}{3}\nabla(\nabla\cdot{\bf u})\right)
\end{eqnarray}
\begin{eqnarray}
\dxdy{T}{t}+({\bf u}\cdot\nabla){T}&=&-v_{r}\left(\frac{d\overline{T}}{dr}-(\gamma-1)\overline{T}h_{\rho}\right)\nonumber\\&  
& {\mbox{}+(\gamma-1)Th_{\rho}u_{r}+\gamma\kappa[\nabla^{2}T+(h_{\rho}+h_{\kappa})\dxdy{T}{r}]}\nonumber\\&
& {\mbox{}+\gamma\kappa[\nabla^{2}\bar{T}+(h_{\rho}+h_{\kappa})\dxdy{\bar{T}}{r}]+\frac{\bar{Q}}{c_{v}}}
\end{eqnarray}
\begin{equation}
\dxdy{{\bf B}}{t}=\nabla\times({\bf u}\times{\bf B})+\eta\nabla^{2}{\bf B}
\end{equation}
Equations 1 and 2 ensure the mass and magnetic flux are conserved. In Equation 3, the momentum equation, g is gravity, $\Omega$ is the rotation rate and C denotes the co-density \citep{bra95,Rogers:2005p1533}, defined as:
\begin{equation}
C=-\frac{1}{\bar{T}}\left(T+\frac{1}{g\bar{\rho}}\dxdy{\bar{T}}{r}P\right)
\end{equation}
where $\bar{T}$ and $\bar{\rho}$ are the reference state temperature and density, and $\partial{\bar{T}}/\partial{r}$ is the reference state temperature gradient.  T, $\rho$ and P are the perturbation temperature, density and pressure, while ${\bf u}$ is the velocity with components $u_{r}$,$u_{\theta}$ and $u_{\phi}$.  

Equation 4 represents the energy equation written as a temperature equation, where $\gamma$ is the adiabatic index (which we specify to be 5/3) and $h_{\rho}$ and $h_{\kappa}$ are the inverse density and thermal diffusivity scale heights, respectively.  As in \cite{Rogers:2005p1533}, we utilize the temperature equation in this formulation of the anelastic equations, as opposed to the entropy formulation commonly used \citep{Glatzmaier:1985p2917}.  This allows us to better specify the super- and subadiabaticity of regions, which can be seen in Equation 4, where the first term on the right hand side (rhs) represents the difference between the reference state temperature gradient and the adiabatic temperature gradient and allows us to represent strongly subadiabatic regions in addition to convection zones.  The last term involving $\bar{Q}$ represents the physics included in the standard solar model, which is not included in this model, that maintains the initial reference state temperature gradient.  The sum of the last two terms in Equation 4 account for the total reference state flux through the system, which is set to zero, so that the reference state temperature is time-independent.  
 
Equation 5 is the magnetic induction equation, in which $\eta$ represents the magnetic diffusivity. 
The thermal diffusivity $\kappa$ is specified to have the solar profile but multiplied by a constant for numerical reasons, as in \cite{Rogers:2005p1533}. The viscous diffusivity is inversely proportional to the density and the magnetic diffusivity is constant:
\begin{equation}
\begin{array}{rrr}
\kappa&=&\kappa_{m}\frac{16\sigma \overline{T}^{3}}{3\overline{\rho}^{2}kc_{p}}\\
\nu&=&\frac{\nu_{m}}{\rho}\\
\eta&=&const\\
\end{array}
\end{equation}
where $\sigma$ is the Stefan-Boltzman constant, k is the opacity and $c_{p}$ is the specific heat at constant pressure.  In the simulations presented here $\kappa_{m}$ is $10^{6}$, $\nu_{m}$ is $ 5 \times 10^{11}$ and $\eta$ is $5 \times 10^{12}$.  Typical values for various parameters are shown in Table 1.  
\begin{table}
\begin{center}
\begin{tabular}{|*{4}{c|}} 
   \hline 
   Parameter & Symbol & Sun & Simulation  \\ 
   \hline
  thermal diffusivity & $\kappa$ & $10^7$ & $1.1 \times 10^{13}$ \\
  magnetic diffusivity & $\eta$ & $10^3$ & $5 \times 10^{12}$ \\
  viscous diffusivity & $\nu$ & $30$ & $2.5 \times 10^{12}$ \\
  \hline  
  Rayleigh number & $gD^{4}\Delta\nabla T_{sup}/T_{bcz}\nu\kappa$ & $10^{23}$ & $10^{6}$\\
  Prandtl number & $\nu/\kappa$ & $3\times 10^{-6}$ & $ 0.22$ \\
  magnetic Prandtl nb & $\nu/\eta$ & $3\times 10^{-2}$ & $ 0.5$ \\
  rotation frequency & $\Omega$ & $ 2.7 \times 10^{-6}$ & $ 2.7 \times 10^{-6}$ \\ 
  Froude number & $(\Omega / N)^2$ & $2 \times 10^{-6}$ & $ 2\times10^{-6}$ \\
  Ekman number & $\!\nu /2\Omega R^2\!$ & $2 \times 10^{-15}$ & $ 2 \times 10^{-4}$ \\
\hline 
\end{tabular}
\bigskip 
  \caption{Values for various parameters in the simulations.  The values listed are those at the base of the convection zone/top of the radiation zone.  Diffusion times are calculated using the depth of the radiation zone and the diffusion values at the base of the convection zone/top of the radiation zone.  The Rayleigh number is calcluated as in \cite{rg05a} and here we calculate the solar value assuming a super-adiabaticity of $10^{-7}$.  Models were run approximately $2 \times 10^{8}s$ and each required approximately a few hundred thousand processor hours.}   
 \label{tble}
   \end{center} 
\end{table}

The method here is similar to that outlined in \cite{Glatzmaier:1984p2768} in which the mass and magnetic flux are written in terms of poloidal and toroidal components:  

\begin{equation}
\overline\rho{\bf u}=\nabla\times\nabla\times(W\hat{r})+\nabla\times(Z\hat{r})
\end{equation}
\begin{equation}
{\bf B}=\nabla\times\nabla\times(B\hat{r})+\nabla\times(J\hat{r})
\end{equation}
The independent variables W, Z, B, J and the thermodynamic variables temperature, T, and pressure P, are then expanded in spherical harmonics so that:

\begin{equation}
\overline{\rho}u_{r}=\frac{1}{r^{2}}\sum_{l} l(l+1)W_{l}(r,t)P_{l}
\end{equation}
\begin{equation}
\overline{\rho}u_{\theta}=\frac{1}{r\sin\theta}\sum_{l} \dxdy{W_{l}(r,t)}{r}\sin\theta\dxdy{P_{l}}{\theta}
\end{equation}
\begin{equation}
\overline{\rho}u_{\phi}=\frac{1}{r\sin\theta}\sum_{l} -Z_{l}(r,t)\sin\theta\dxdy{P_{l}}{\theta}
\end{equation}
and similarly for the magnetic field
\begin{equation}
B_{r}=\frac{1}{r^2}\sum_{l} l(l+1)B_{l}(r,t)P_{l}
\end{equation}
\begin{equation}
B_{\theta}=\frac{1}{r\sin\theta}\sum_{l} \dxdy{B_{l}(r,t)}{r}\sin\theta\dxdy{P_{l}}{\theta}
\end{equation}
\begin{equation}
B_{\phi}=\frac{1}{r\sin\theta}\sum_{l} -J_{l}(r,t)\sin\theta\dxdy{P_{l}}{\theta}
\end{equation}
where $\theta$ is the co-latitude, ${l}$ is the spherical harmonic degree and $P_{l}$ are the Legendre polynomials.  Temperature and pressure are similarly expanded so that: 
\begin{equation}
T=\sum_{l}T_{l}(r,t)P_{l}
\end{equation}
\begin{equation}
P=\sum_{l}\mathcal{P}_{l}(r,t)P_{l}
\end{equation}
The radial functions $W_{l}(r,t)$, $Z_{l}(r,t)$, $B_{l}(r,t)$, $J_{l}(r,t)$,  $P_{l}(r,t)$ and $T_{l}(r,t)$ are discretized on a non-uniform, second-order finite difference grid in radius.  This is done to allow us the flexibility of enhanced resolution around the tachocline and overshoot layer, without having the burden of insufficient resolution at the tachocline, required by the Chebyshev grid or alternatively, enhanced resolution at the bottom of the radiation zone (as would be necessary with stacked Chebyshev grids).  

We then solve the radial component of the momentum equation for $W_{l}$, the radial component of the curl of the momentum equation for $Z_{l}$ and the full divergence of the momentum equation for $P_{l}$.  The equation for $P_{l}$ is then a diagnostic equation, unlike the Glatzmaier code \cite{Glatzmaier:1984p2768} which solve the horizontal divergence of the momentum equation.  This is done to avoid third order derivatives, which would complicate boundary conditions when using the finite difference scheme employed here.  We impose the stress-free and impermeable boundary conditions on the velocity.

For the magnetic terms we solve the radial component of the magnetic induction equation for $B_{l}$ and the radial component of the curl of the magnetic induction equation for $J_{l}$.  The temperature and pressure equations are solved directly for $T_{l}$ and $P_{l}$.  The boundary conditions on temperature are constant temperature perturbation, on velocity we force stress-free and impermeable conditions, the magnetic field is matched to an internal potential field at the bottom of the domain and an external potential field at the top of the domain.  The current density, $J_{l}$ is zero on the top and bottom boundary.

Timestepping is accomplished by an explicit Adams-Bashforth method for the non-linear terms and an implicit Crank-Nicolson method for the linear terms.  The code is parallelized using Message-Passing-Interface (MPI).  Each model was run approximately $1-2\times 10^{8}s$ and took approximately a few hundred thousand processor hours.    

\clearpage

\clearpage

\clearpage
\clearpage
\clearpage
\clearpage

\clearpage

\clearpage

\end{document}